\documentclass[12pt,titlepage]{article}
\usepackage[utf8]{inputenc}
\usepackage[margin=1in,letterpaper]{geometry}
\usepackage[nodisplayskipstretch]{setspace}

\usepackage{mathtools,amssymb,microtype,bm,dsfont,graphicx,color,soul,placeins,booktabs,multicol,multirow,rotating,natbib,etoolbox}
\usepackage{adjustbox,bbm,listings}
\usepackage{titling}\thanksmarkseries{fnsymbol}
\pretitle{\begin{center}\Large}
\preauthor{\begin{center}\normalsize\lineskip 1em \begin{tabular}[t]{c}}
\predate{\begin{center}\normalsize}

\definecolor{green}{rgb}{0.0,0.47,0.44}
\usepackage{marginnote}
\newcounter{ctrpend}

\bibpunct[, ]{(}{)}{,}{a}{}{,}

\setcitestyle{notesep={:}}\newcommand{\p}[1]{#1}\newcommand{\pp}[1]{#1}

\usepackage{titlesec}

\let\oldappendix\appendix
\renewcommand{\appendix}{\oldappendix\titleformat{\section}[block]{\normalfont\Large\bfseries}{Appendix~\thesection:~}{0em}{}\setcounter{figure}{0}\numberwithin{figure}{section}\setcounter{table}{0}\numberwithin{table}{section}}

\usepackage[breaklinks,hidelinks]{hyperref}

\newcommand{\one}{\mathbbm{1}}
\newcommand{\ud}{\,\mathrm{d}}
\newcommand{\Keywords}{\bigskip\noindent\textbf{Keywords:\ }}
\newcommand{\E}{\mathbb{E}}
\renewcommand{\P}{\mathbb{P}}

\AtBeginEnvironment{abstract}{\onehalfspacing}

\newcommand{\TITLE}{Revisiting the Unitary Actor Assumption:\\Toward Realistic Aggregation of Individual Preferences in Strategy Research}
\title{\TITLE\thanks{We thank Senior Editor Tobias Kretschmer and three anonymous reviewers for their guidance and constructive feedback throughout the review process.  We are also grateful to Daniel Albert, Dylan Boynton, Javier Gimeno, Tomasz Obloj, Kramer Quist, and Brit Sharoni for their insightful comments on earlier drafts.  Finally, we thank seminar participants at Wharton, ESSEC, Emory, UT Austin, Harvard Business School, the SMS Annual Conference, and the Academy of Management Meeting for their helpful suggestions.  All remaining errors are our own.}}
\author{Felipe A. Csaszar \\ Ross School of Business \\ University of Michigan \\ Ann Arbor, MI 48109 \\ \texttt{fcsaszar@umich.edu} \and John C. Eklund \\ Marshall School of Business \\ University of Southern California \\ Los Angeles, CA 90089 \\ \texttt{jceklund@marshall.usc.edu}}
\date{{\vskip 2em}\today}
\date{{\vskip 1em}\today{\vskip 2em}Forthcoming in \emph{Strategy Science}}

\begin{document}
\begin{onehalfspace}\maketitle

\begin{abstract}
The long-standing unitary-actor assumption in strategy research---treating firms as monolithic entities with coherent preferences---misses that organizations are coalitions of individuals with diverse and often conflicting goals.  Although behavioral perspectives have challenged this assumption, the field lacks an operational method for deriving an organizational utility function from the disparate preferences of its members and the specific structures used to aggregate them.  We develop a mathematical framework that (i) maps individual utility functions into choice probabilities via a random-utility model, (ii) combines those probabilities using an explicit aggregation structure (e.g., unanimity or polyarchy), and (iii) recovers an organizational utility function that rationalizes the collective behavior.  This establishes organizational utility functions as operationally meaningful: they summarize and predict organizational choice, yet are generally not simple averages of members' utilities.  Instead, aggregation structures systematically reshape preferences---unanimity approximates the pointwise minima of underlying utility functions, amplifying risk aversion; polyarchy approximates the pointwise maxima, promoting risk-seeking.  We illustrate strategic implications in Cournot competition and principal-agent settings, showing how internal aggregation structures shift competitive and collaborative outcomes.  Overall, the framework provides a parsimonious way to retrofit unitary-actor models with behaviorally grounded organizational preferences, reconciling the coalition view of the firm with rigorous and tractable strategic analysis.

\Keywords aggregation; utility; organizational structure; group decision-making; evaluation
\end{abstract}
\end{onehalfspace}

\section{Introduction}

\subsection{Questioning the Unitary Actor Assumption}

Strategy research is fundamentally concerned with elucidating the factors that drive performance heterogeneity among organizations \citep{Nag07,Leib18}.  A critical, yet often implicit, assumption underpinning many theories that seek to explain performance heterogeneity---such as those found in industrial organization \citep{Tiro88} and search models of firms \citep{Levi97}---is the conceptualization of organizations as \emph{unitary actors}.  For instance, industrial organization scholars typically depict firms as profit-maximizing, risk-neutral entities, whereas the literature on search typically depicts companies as singular entities navigating complex business landscapes in pursuit of high-fitness positions.  However, the unitary actor assumption has been increasingly challenged by insights from the behavioral theory of the firm and the more recent literature on microfoundations of strategy \citep[see, e.g.,][]{Cyer63,Feli15}.  These bodies of work underscore that organizations are not unitary actors but are comprised of diverse actors with potentially conflicting micro-level preferences, leading to varied and complex organizational behaviors at the macro level.

The literature on aggregation has made significant strides in bridging the gap between micro and macro levels \citep{Knud07,Csas13}.  This work has sought to infer organizational decisions by combining two elements: individuals' choice probabilities and the organization's aggregation structure.  By combining these elements, researchers can predict how individual decisions coalesce into organizational outcomes.  For example, if two individuals each have a 50\% chance of approving a project, the probability that an organization using unanimity decision-making will approve it is only 25\%.  This basic example illustrates how aggregation can lead to more conservative organizational behavior compared to individual tendencies.  By explicitly modeling these processes, researchers can better understand the emergent properties of organizational behavior, offering valuable insights into the complex dynamics that shape firm-level outcomes and performance heterogeneity (we review some of this work in Section~\ref{sec:agg-probs}).

Despite these advances, understanding aggregation through choice probabilities is not fully satisfactory, as utility functions---not choice probabilities---are the central construct in decision-making research.  Utility functions are the \emph{lingua franca} of this field, as they help consistently specify decision-makers' preferences, offer rich insights into characteristics such as risk aversion and preference strength, and are supported by robust empirical methods like discrete-choice analysis \citep{McFa74}.  Utility functions are crucial for understanding decisions at both the individual level (e.g., prospect theory research; \citealt{Kahn79}) and the firm level (e.g., the industrial organization and search literatures mentioned above).  However, research has largely neglected how utility functions aggregate, with few exceptions assuming that group utility functions average individual utilities \citep{Adam05,Marc14}; yet, it is unclear what actual aggregation process, if any, corresponds to averaging individual utilities.

Thus, a fuller treatment of aggregation calls for examining how an organization's utility function is derived from the utility functions of its members and the aggregation structure employed.  Currently, there is no robust theoretical foundation for deriving organizational utility functions, raising critical questions about the objectivity of aggregating individual utility functions, and the accuracy of such functions in reflecting an organization's overall preferences.  At stake is the validity of the unitary actor assumption upon which much of strategy research relies.

To address these challenges, we pose the following research question: How do organizational utility functions emerge from individual preferences and aggregation structures?  By exploring this question, we aim to develop a theoretical framework capable of translating individual-level preferences into organizational utility functions, thereby providing a more nuanced perspective on organizational behavior and strategic decision-making.

\subsection{Our Approach and Contribution}

To answer our research question, we introduce a mathematical method to derive an organization's utility function given its members' utility functions and the aggregation structure they use.  The key insight we use to derive the organizational utility function is to turn individuals' utility functions into choice probabilities (i.e., screening functions); we then aggregate these probabilities using the relevant aggregation structure, and determine what organizational utility function would produce such aggregate choice probabilities (i.e., the utility function that rationalizes the organization's behavior).  Our methodology is flexible and can accommodate arbitrarily complex utility functions, any number of individual actors, and aggregation structures.  For clarity, the bulk of our analyses illustrate the simplest possible aggregation structures and individuals' utility functions.  In terms of structures, we study unanimity and polyarchy.  Organizations using unanimity only approve projects if all members approve; organizations using polyarchy approve projects if any single member approves \citep{Sah86}.  In terms of individuals' utility functions, we focus on single-variable linear functions.  Most of our analyses focus on two-member organizations.  Once the mechanisms underlying these basic cases are understood, we illustrate how to include non-linear utility functions, utility functions that depend on multiple variables, and more than two decision makers.

By developing this theoretical framework, we make several interconnected contributions to the strategy literature.  First, we demonstrate that organizational utility functions are meaningful constructs that often differ substantially from those of the underlying individuals, and that the method of aggregation fundamentally shapes these functions.  This advances our understanding of organizational decision-making by providing analytical tools to predict systematic biases in organizational choices.  Second, our approach enables the ``retrofitting'' of unitary actor theories of the firm with behaviorally plausible organizational utility functions, establishing more realistic foundations for these models.  Third, we contribute to one of the ``forgotten pillars'' of the behavioral theory of the firm \citep[\p 528]{Gave07}---conflict within organizations---by developing a new understanding of how conflicting preferences aggregate.  Specifically, our framework resolves the apparent contradiction highlighted by \citet[\p 27]{Cyer63}, who noted that ``the idea of an organization goal and the conception of an organization as a coalition are implicitly contradictory.''  Additionally, our approach tackles a long-standing issue in organizational theory identified by \citet[\pp 668--671]{Marc62} by providing a method that is both analytically tractable and capable of representing internal organizational conflicts.

The method we outline illuminates various strategy-related phenomena by revealing how decision-making bodies like top management teams arrive at organizational choices.  These teams often encompass individuals with diverse preferences \citep{Hamb84,Fink09} and who may belong to conflicting coalitions \citep{Mith20,Zhan19}.  Understanding this aggregation process offers insight into how senior managers' attributes and organizational decision structures influence strategic outcomes, such as firm responses to technological change \citep{Egge18}, the aggressiveness of acquisition strategies \citep{Hale09}, and overall organizational risk preferences \citep{Lim14}.  Practically, our framework provides guidance for organizational design, helping leaders structure decision-making processes to better align with strategic objectives.

The paper is structured as follows.  The next section reviews prior studies that have leveraged the unitary actor assumption and existing research on aggregation.  We then derive our model, present illustrative results, and conclude by developing broader theoretical and practical implications of our work.

\section{Theoretical Background}

This section situates our research within the context of previous studies.  First, we illustrate how common the unitary actor assumption is in strategy research.  Next, we explore the limitations associated with this assumption.  Finally, we examine existing studies on aggregation.

\subsection{The Unitary Actor Assumption in Strategy Research}

Strategy scholars frequently employ the unitary actor assumption to simplify the theoretical treatment of firms.  This assumption posits that an organization can be conceptualized as a single entity with a coherent set of preferences represented by a utility function.  From a decision-making perspective, this approach treats a firm, no matter how large or complex, as an individual actor striving to achieve specific goals.

The unitary actor assumption is prevalent across various theoretical frameworks in strategy research.  Prominent examples include the structure-conduct-performance paradigm, the resource-based view (RBV), positioning, evolutionary economics, Schumpeterian competition, and most game-theoretic models of competition.  This assumption is evident in statements that attribute decision-making to ``the firm,'' such as ``a firm \emph{seeks} to be unique in its industry'' \citep[\p 16]{Port85}, ``the consuming firm \emph{decides} to make this product'' \citep[\p 13]{Coas88}, or ``the firm \emph{seeks} to enter an emerging market'' \citep[\p 415]{Tusc13} (emphases added).

Economic models of competition predominantly adopt this assumption, typically positing that firms have a singular objective: profit maximization (often coupled with risk-neutrality).  The unitary actor assumption is evident in industrial organization models, spanning from \citet{Cour38} to contemporary research.  Indeed, with the notable exception of principal--agent models, all models presented in standard introductions to industrial organization \citep[e.g.,][]{Tiro88} employ the unitary actor conceptualization of firms.

The prevalence of this assumption extends beyond economics to encompass behavioral theories in strategy.  For instance, the literature on organizational search often depicts the firm as a unitary actor navigating a rugged NK landscape \citep{Levi97}.  In this literature, firms are portrayed as making decisions regarding which position on the landscape to explore next, with the aim of enhancing overall performance.  Similar unitary actor representations are used in other behavioral theories of the firm, including those addressing learning, routines, attention, and aspirations.

The widespread adoption of the unitary actor assumption in strategy research is convenient, as it facilitates theoretical parsimony and enables scholars to analyze firm behavior as that of a homogeneous entity pursuing well-defined objectives.  However, researchers have questioned whether this simplifying assumption can accurately reflect the complexity of real organizations.  The next section critically examines this assumption.

\subsection{Challenging the Unitary Actor Assumption}

The unitary actor assumption has been challenged by three major streams of research: the behavioral theory of the firm, the microfoundations of strategy, and the economics of organizations.

In their seminal work, \citet{Cyer63} challenge the notion that firms operate as unitary actors by emphasizing the role of conflicting interests within the firm.  They argue that decision-making within firms is not centralized or monolithic but rather the result of a complex process of negotiation, compromise, and conflict resolution among various internal groups, such as managers, employees, shareholders, and other stakeholders.  Cyert and March's behavioral theory highlights the importance of understanding the internal dynamics and decision-making processes within firms to fully grasp firm behavior.  The unitary actor assumption, which is equivalent to viewing the firm as a black box, overlooks these crucial internal processes and conflicts.

Although conflict played a central role in Cyert and March's work, subsequent research has ``largely overlook[ed] the potential for conflict in decision-making'' \citep[\p 268]{Jose20}.  Similarly, \citet[\p 528]{Gave07} describe the role of conflict within firms as a ``forgotten pillar'' of the Carnegie tradition.

More recently, and in a similar vein to the ideas inspiring the behavioral theory of the firm, the literature on the microfoundations of strategy has highlighted that to understand firm performance, one must look within the firm, at the individuals and processes occurring within it \citep{Barn13,Feli15}.  Studies in this domain have linked characteristics of the individuals making firm decisions to firm performance.  Characteristics studied include managers' cognitive capabilities \citep{Helf15}, individuals' skills and knowledge \citep{Loon20}, managers' heuristics \citep{Bing19}, acquisition experience \citep{Meye19}, scientific expertise \citep{Hess11}, and understanding of organizational routines \citep{Aime09}.

Despite the intense focus on individual characteristics, the microfoundations literature has paid less attention to how these characteristics are aggregated to shape organization-level behaviors and outcomes \citep[\pp 6--7]{Cowe22}.  Understanding this aggregation process is crucial for a comprehensive view of firm behavior, as it bridges the gap between individual actions and collective organizational outcomes \citep{Engl19}.

Parallel to these behavioral and microfoundational perspectives, the economics of organizations has also formalized departures from the unitary actor assumption.  Seminal work by \citet{Coas37} and \citet{Will75} opened the black box of the firm, viewing it as a governance structure designed to mitigate transaction costs arising from bounded rationality and opportunism.  Agency theory further formalized the divergence of interests between owners and managers \citep{Jens76,Holm79}, while property rights theory emphasized how incomplete contracts shape power dynamics between distinct parties \citep{Gros86,Hart95}.  These theories---recognized by multiple Nobel Memorial Prizes in Economic Sciences---explicitly model the firm as composed of distinct actors with disparate utility functions.  However, while organizational economics rigorously analyzes conflict and the mechanisms (contracts, incentives, authority) used to align interests, it typically does not seek to derive a single aggregate utility function for the organization itself.  Instead, it focuses on equilibrium outcomes of strategic interactions between members.

Despite these theoretical advances across behavioral, microfoundational, and economic traditions, strategy research still lacks an operational concept of organizational utility that accounts for these internal complexities.  While organizational economics focuses on equilibrium outcomes of strategic interactions, and behavioral scholars focus on process, neither stream typically derives a single aggregate function that can represent the firm's collective preferences.  Consequently, researchers often revert to the simplistic unitary actor assumption for tractability.  The rest of this section examines two different approaches that have been used to address this aggregation issue.  First, scholars have sought to aggregate individual utility functions.  Second, other studies have focused on aggregating the probabilities of individuals selecting a project to capture the preferences of the overall organization.  Each approach has its own unique set of strengths and weaknesses that we outline below.

\subsection{Approach 1: Aggregating Utility Functions}

This approach has been used within the marketing literature to understand the consumer behavior of groups (e.g., a family deciding on a vacation or a committee deciding on which supplies to acquire; see, e.g., \citealt{Aror99,Mena89}).  This research posits that the organizational utility function resembles the average of the utility functions of the constituent individuals.  Sometimes the average is replaced by a weighted average to account for situations where some of the members (e.g., the parents) have more power than other members (e.g., the children).

Although this approach does not view the organization as a black box, it shares a similar drawback with the unitary actor approach: the realism of its organizational utility function is questionable.  Specifically, it \emph{assumes} that aggregation averages individuals' utility functions.  This assumption of a simple averaging function is not necessarily accurate, as the approach does not \emph{derive} the organizational utility function from the individuals' utility functions and the actual aggregation mechanism they use.  Moreover, we know that the aggregation mechanism an organization uses changes its organizational choices \citep{Csas12}, so it must inevitably also alter the organizational utility function.  In contrast, simply averaging utility functions does not account for the specific aggregation mechanism used.

Some research has taken a ``negative'' approach toward the question of aggregating utility functions.  Two objections have been raised when discussing the idea of organizational utility functions: that interpersonal utility comparisons are intractable \citep{Hamm91,Binm09}, and that it is impossible to aggregate preferences without violating seemingly simple and reasonable conditions (i.e., Arrow's \citeyear{Arro51} ``impossibility theorem'').

However, these difficulties need not apply to the specific problem of aggregating preferences within organizations.  As (i) one can derive organizational utility functions without making interpersonal utility comparisons (e.g., rather than aggregating utilities, the first stage of our method aggregates choices) and (ii) Arrow's \citeyearpar{Arro51} theorem is about the impossibility of an \emph{optimal} aggregation structure (where optimality is defined in terms of satisfying a basic set of reasonable conditions: unrestricted domain, independence of irrelevant alternatives, weak Pareto principle, and non-dictatorship), whereas we just seek to describe the utility function corresponding to an \emph{actual} aggregation structure such as unanimity or polyarchy (i.e., optimality is not a requirement).

In sum, the existing approach to deriving organizational utility functions is limited in the sense that it makes the ad-hoc assumption that the organizational utility function must be an average of individuals' utility functions.  We now turn to a second approach toward aggregation, which avoids this ad-hoc assumption at the expense of not deriving utilities but choice probabilities.

\subsection{Approach 2: Aggregating Choice Probabilities}\label{sec:agg-probs}

Because the approach of aggregating individual utility functions has significant limitations, scholars have sought to circumvent these problems by looking at how individual \emph{choices} (rather than utilities) are aggregated.  This approach stems from social choice theory's use of voting rules \citep{Blac58,Cond1785} and was formalized by \citet{Davi73} through Social Decision Schemes (SDS).  SDS provide a mathematical framework where individual choice probabilities across alternatives are aggregated via rules like majority, proportionality, or unanimity to yield a group probability distribution.  Subsequent work explored when aggregation improves judgments over single experts \citep{Hoga78,Clem89}, noting the effectiveness of simple rules like majority \citep{Hast05}, and examined how individuals use others' opinions \citep{Soll09} or how opinions evolve socially \citep{Frie11}.

Distinct from, yet related to, SDS is the literature on decision-making structures initiated by \citet{Sah86} and extended by others \citep[e.g.,][]{Knud07,Chri10,Csas12,Csas13}.  This literature examines how hierarchies and polyarchies aggregate individual members' probabilities of accepting opportunities based on perceived quality, shifting the focus from social choice rules to organizational design and how structures influence evaluation quality under uncertainty.  Like SDS, this approach operates in probability space rather than utility space; for instance, if two members each have a 50\% chance of approving a project, a unanimity structure yields only a 25\% approval probability.

\begin{figure}\centering
\includegraphics[width=\linewidth]{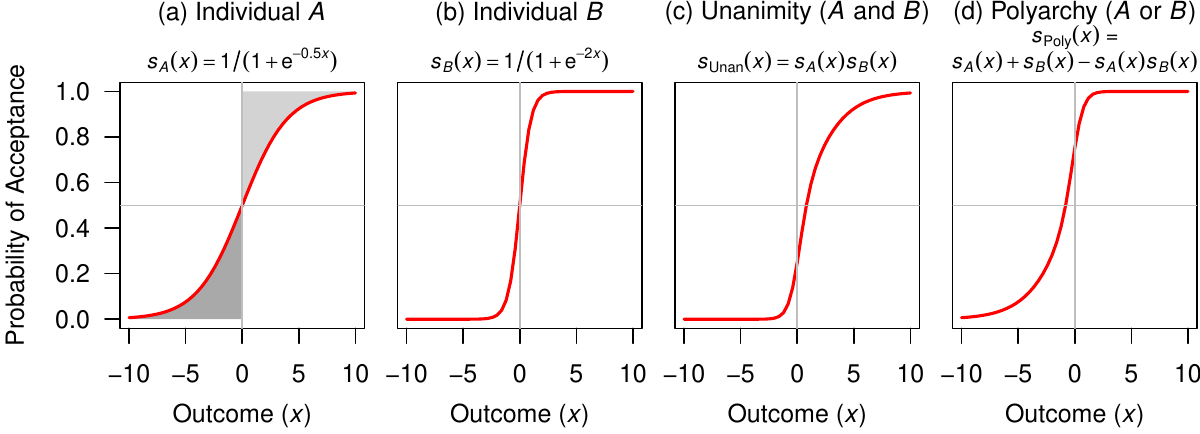}
\caption{Examples of screening functions.}\label{fig:ex-screening}
\end{figure}

Instead of using utility functions, this approach relies on the concept of a ``screening function,'' a function that maps the outcome of a project to the probability of that project being accepted.  The simplest screening function would be one that approves all projects that have a positive outcome and rejects all projects with negative outcomes (i.e., this is a step function whose probability of approval is $s(x) = \one[x>0]$); however, because screening by fallible individuals is imperfect, the probability of approval may only increase slowly as outcomes shift from negative to positive.  Figure~\ref{fig:ex-screening} illustrates four different screening functions.

Panel (a) shows a typical screening function, one where the probability of accepting a project increases smoothly as the outcome ($x$) of a project improves.  The $x$-axis of a screening function is the same as the $x$-axis in a utility function: the outcome being assessed (e.g., the quality of a project).  Note that for the screening function in panel (a), a project with a negative outcome (e.g., $x=-1$) may be accepted and one with a positive outcome may be rejected.  As previously mentioned, this happens because individuals are fallible: they do not see the actual outcome of a project but a noisy perception of it.  The area to the right of $x=0$ and above the screening function (shaded in light gray in panel (a)) determines the probability of making an omission error (i.e., rejecting a good project).  Similarly, the area to the left of $x=0$ and below the screening function (shaded in dark gray) determines the probability of making a commission error; that is, accepting a bad project.

Panel (b) illustrates a much sharper screening function, one in which the probability of accepting projects increases steeply around $x=0$.  The individual described in panel (b) is more accurate than the individual described in panel (a), as she is less likely to make either omission or commission errors.

Panel (c) illustrates a screening function that is shifted farther to the right than those previously described.  This shift to the right means that a project needs to have a substantially better outcome to have a good chance of being accepted.  Conversely, the screening function in panel (d) is shifted to the left and, hence, even projects with negative outcomes have a high chance of being approved.

One can aggregate individual-level screening functions into an organization-level screening function by using basic probability theory.  For instance, an organization with two members ($A$ and $B$) that approves when $A$ \textsc{and} $B$ approve has a screening function corresponding to the multiplication of the members' screening functions.  So, if $A$ and $B$ have the screening function depicted in panels (a) and (b) respectively ($s_A(x) = \frac{1}{1 + e^{-0.5x}}$ and $s_B(x) = \frac{1}{1 + e^{-2x}}$), then unanimity's screening function would be $s_\text{Unan}(x) = s_A(x) s_B(x) = \frac{1}{(1 + e^{-0.5 x})(1 + e^{-2 x})}$.  Similarly, an organization that approves when $A$ \textsc{or} $B$ approve has a screening function $s_\text{Poly}(x) = s_A(x) + s_B(x) - s_A(x) s_B(x) = 1 - \frac{1}{(1 + e^{0.5 x})(1 + e^{2 x})}$.

The customary name for an organization that approves when at least one of its members approves is \emph{polyarchy} and the customary name for an organization that only approves when everybody approves is \emph{unanimity} \citep[or ``hierarchy'';][]{Sah86,Sah88,Csas12}.  Because unanimity requires everybody to agree, unanimity is a more conservative rule than polyarchy.  Graphically, this means that the screening function of unanimity is shifted to the right vis-à-vis polyarchy (i.e., unanimity will require a higher $x$ than polyarchy to approve projects).  In fact, the curves in panels (c) and (d) are respectively the screening functions of unanimity and polyarchy derived above.  \citet{Sah88} extend these formulae to $N$ individuals as well as to cases where the number of approvals required can be any number between 1 and $N$.  \citet{Csas13} analyzes an even larger set of structures and \citet{Csas12} tests the predictions of \citet{Sah86} using data from mutual funds.

The main benefit of using screening functions is that it is possible to derive an organization's behavior from an understanding of its members and the aggregation structure used.  This allows for determining useful metrics such as the probability that an organization will make omission and commission errors.

An important limitation of the screening function approach is that it does not say anything about the organization-level utility function---a key drawback given that utility functions are central to decision-making research.  They offer critical insights into decision-makers' characteristics (e.g., risk preferences) and are essential for accurately reflecting an organization's overall preferences, which is crucial for validating the unitary actor assumption in strategy research.

Another limitation is that unlike utility functions, which have a clear connection with individual preferences, it is unclear how screening functions represent preferences.  This makes it difficult to use screening functions to examine issues related to conflict and incentives within organizations.  In fact, the research on screening functions has generally avoided problems related to incentives and conflict in organizations \citep[\p 268]{Jose20} and instead adhered to the ``team-theoretic assumption'' \citep{Mars72}.

In sum, despite several potential benefits, prior research has not provided a way to derive an organization's utility function from the utility functions of its members and the aggregation structure they use.  The next section introduces a way of doing so.

\section{Method}\label{sec:method}

\subsection{Overview of the Method}

The key insight underlying our method is that one can avoid the difficulties of aggregating utilities by turning individual utility functions into individual screening functions, then aggregating these screening functions, and finally turning the aggregate screening function into an aggregate utility function.  In shorthand, our approach moves from utility-space to probability-space, performs the aggregation there, and then comes back to utility-space.

\begin{figure}\centering
\includegraphics[width=\linewidth]{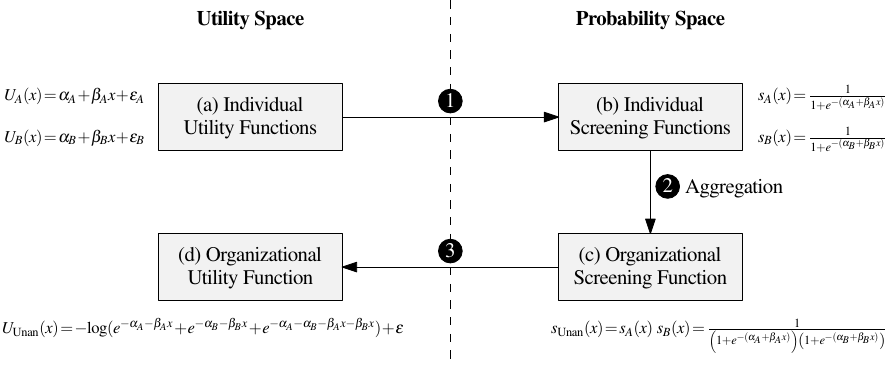}
\caption{Summary of the method used to go from individuals' utility functions to an organizational one.}\label{fig:functions-diagram}
\end{figure}

More specifically, our approach has three steps: (1) convert the utility functions of the members of the organization into screening functions; (2) aggregate the screening functions using the method described in the previous subsection; and (3) convert the aggregate screening function into an aggregate utility function, which we call the \emph{organizational utility function}.  This last step involves looking at the organization's behavior and asking ``if this was the behavior of a person, what would be the utility function that would explain this behavior?''  These three steps are illustrated as the arrows in Figure~\ref{fig:functions-diagram}, which presents an overview of the main elements (for now, disregard the equations in the figure, which are explained later).

Following the literature on discrete choice \citep{McFa74}, our model assumes that utility (both individual and organizational) includes a random term that captures decision-making inconsistencies due to perceptual and cognitive limitations.  For instance, the utility of being paid $\$x$ could be $U(x) = x^{0.5} + \varepsilon$, where $\varepsilon$ is a random variable.\footnote{The random utility assumption originated in early psychophysics research (\citealt{Fech1860,Thur27}; see also Chapter~2 in \citealt{Luce59}) and now forms the foundation of the empirical research on decision-making (for a survey, see \citealt{Hess14}; for a recent popular treatment, see \citealt{Kahn21}).  \citet{Mans77} identifies two distinct interpretations of the random error: psychological (representing inherent noise in human cognitive processes) and econometric (capturing researchers' incomplete information about alternatives or decision makers).  Our paper specifically addresses noisy evaluation, making the psychological interpretation most relevant to our analysis.}  The customary assumption regarding this random distribution is to use a $\mathrm{Logistic}(0,1)$, similar in shape to the Normal distribution but analytically tractable.  Note that random utility models typically keep this distribution fixed.  This assumption does not make random utility models less general, as scaling the other terms of the utility function has the same effect as scaling the random term.

We say that our model is derived from first principles because it rests on two basic assumptions: (i) that individuals can be modeled using a random-utility framework, and (ii) that organizational aggregation can be described as a rule that combines members' choice probabilities into an organizational choice probability.  As shown in the previous section, both assumptions are well-established in the decision-making literature.  Because our derivation depends only on these general assumptions---rather than on specific functional forms---our framework can accommodate any form of individual utility function and any aggregation structure.

Mathematically, this generality can be expressed as follows: let $F$ denote the function that converts utilities into probabilities (arrow 1 in Figure~\ref{fig:functions-diagram}), $F^{-1}$ its inverse (arrow 3), and $G$ the function that aggregates probabilities according to the organization's structure (arrow 2).  Then the utility function of an organization with any number of members and any aggregation structure is simply $F^{-1}\left(G\left(F\left(u_1(x)\right), \ldots, F\left(u_N(x)\right)\right)\right)$.  This formulation extends our method to any situation where these functions can be defined.

For clarity of exposition, we illustrate our approach using simple utility functions and aggregation structures.  Specifically, we analyze single-variable linear utility functions with positive slopes, representing situations where ``more is better'' (i.e., $U(x) = \alpha + \beta x + \varepsilon$, with $\beta > 0$), and two two-member aggregation structures: unanimity and polyarchy.  These simplifying choices yield tractable results and clear insights into the determinants of organizational utility functions; they can be interpreted as representing small groups like top management teams.

Later in the paper we relax these assumptions in several directions: Section~\ref{sec:larger-n} considers larger groups ($N>2$), Section~\ref{sec:opposing-views} allows members with opposing preferences, Section~\ref{sec:strat-interactions} embeds the resulting organizational utility functions in competitive (Cournot) and collaborative (principal--agent) settings, and Appendix~\ref{app:complex-utfns} explores more complex individual utility functions.

We now derive the utility functions for unanimity and polyarchy.  Because much of the setup is common to both, we present most technical details in the first derivation.

\subsection{Unanimity}

We assume an organization with two members, $A$ and $B$, whose utility functions are:
\begin{align*}
U_A(x) &= \alpha_A + \beta_A x + \varepsilon_A \\
U_B(x) &= \alpha_B + \beta_B x + \varepsilon_B
\end{align*}

That is, the $\beta$'s control how sensitive each individual is to the value $x$ of the project being evaluated, and the $\alpha$'s control their baseline level of satisfaction (e.g., an individual with a high $\alpha$ is akin to an optimist, someone who tends to like most projects; and vice versa for low $\alpha$).  The $\varepsilon$'s are random draws from a logistic distribution with mean 0 and scale parameter~1.  We assume members' evaluation errors are independent; relaxing this is straightforward but changes the aggregation algebra.

\begin{figure}\centering
\includegraphics[width=\linewidth]{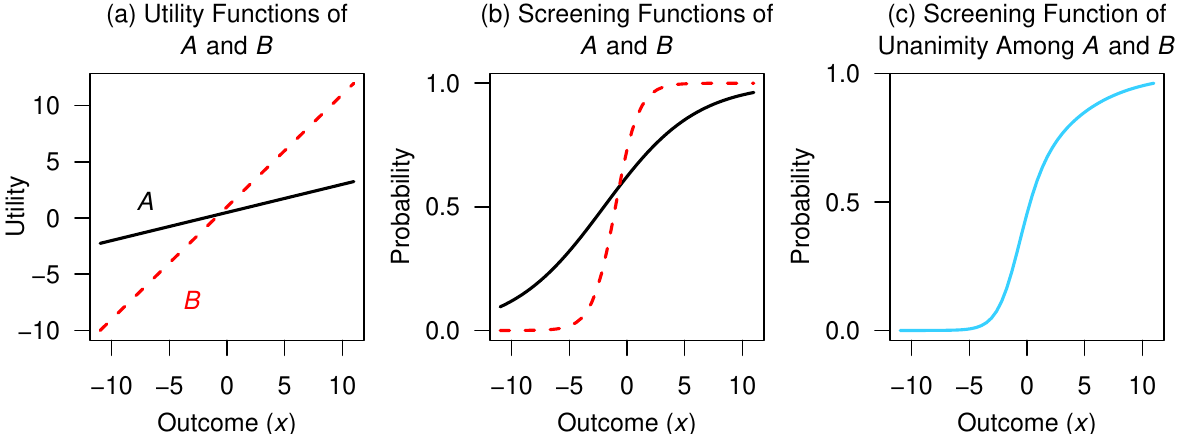}
\caption{Illustration of the initial steps of the method: going from (a) individual utility functions to (b) individual screening functions to (c) an organizational screening function.}\label{fig:ex-utility-screening}
\end{figure}

For the sake of clarity, we illustrate the model with a numerical example.  We start this example with panel (a) in Figure~\ref{fig:ex-utility-screening}, which plots the nonrandom part of the utility functions of two individuals, one with utility $U_A(x) = 0.5 + 0.25x + \varepsilon_A$ and the other with $U_B(x) = 1 + x + \varepsilon_B$.  In this example, individual $B$ is both more optimistic and more sensitive than individual~$A$.

The screening functions of these individuals---the probabilities that each recommends approval---are the probabilities that, given a project with outcome $x$, each individual's utility is positive.  That is,
{\small
\begin{align*}
s_A(x) &= \P[U_A(x) > 0] = \int_{-\infty}^{\infty} f(\varepsilon_A) \one[\alpha_A + \beta_A x + \varepsilon_A > 0] \ud \varepsilon_A = F(\alpha_A + \beta_A x ) = \frac{1}{1 + e^{-(\alpha_A + \beta_A x)}} \\
s_B(x) &= \P[U_B(x) > 0] = \int_{-\infty}^{\infty} f(\varepsilon_B) \one[\alpha_B + \beta_B x + \varepsilon_B > 0] \ud \varepsilon_B = F(\alpha_B + \beta_B x ) = \frac{1}{1 + e^{-(\alpha_B + \beta_B x)}},
\end{align*}
}\noindent
where $f(\cdot)$ and $F(\cdot)$ are, respectively, the probability density and cumulative distribution of the logistic distribution.  Thus, the screening functions of the individuals in the running example are $s_A(x) = \frac{1}{1+e^{-(0.5+0.25x)}}$ and $s_B(x) = \frac{1}{1+e^{-(1+x)}}$.  Panel (b) in Figure~\ref{fig:ex-utility-screening} plots the screening functions of these individuals.

Now that we have expressions for the screening functions of both individuals, we can compute the screening function of the organization.  That is, the probability that the organization would approve a project with a given outcome $x$.  Because unanimity approves a project when both individuals recommend approving it, the screening function of the organization is
\begin{equation}\label{eq:prob-unan}
s_{\text{Unan}}(x) = s_A(x)\ s_B(x) = \frac{1}{( 1 + e^{- ( \alpha_A + \beta_A x )} )( 1 + e^{- ( \alpha_B + \beta_B x )} )}.
\end{equation}
Panel (c) in Figure~\ref{fig:ex-utility-screening} adds this probability to the ongoing example.

We are now in the position to ask what is the utility function that would generate the screening function of unanimity.  To make this utility function comparable to those of individuals, we assume it also includes a logistic error term: $U_{\text{Unan}}(x) = u_{\text{Unan}}(x) + \varepsilon$, where $\varepsilon \sim \mathrm{Logistic}(0,1)$.\footnote{The error term in the organizational utility function can be interpreted as a form of ``cognitive'' error at the organizational level---that is, inconsistency or noise in the organization's utility function and choices.  This error term fully captures the inconsistency in the organization's decision-making.  Under the $\mathrm{Logistic}(0,1)$ normalization for the random-utility errors (which fixes the scale), the mapping from acceptance probabilities to deterministic utility is unique: $u(x)=\text{logit}(s(x))=\log\left(s(x) / \left(1-s(x)\right)\right)$.  Consequently, arrow~3 in Figure~\ref{fig:functions-diagram} is reversible in the same way that arrow~1 is reversible for each individual.}  We know that someone with such a utility function would approve projects with probability
\begin{equation}\label{eq:prob-pi}
s(x) = \P[U_{\text{Unan}}(x) > 0] = \frac{1}{1 + e^{-u_{\text{Unan}}(x)}}.
\end{equation}

Now we solve for the $u_{\text{Unan}}(x)$ that makes equations \eqref{eq:prob-unan} and \eqref{eq:prob-pi} equal.  That is,
\begin{align*}
s(x) &= s_{\text{Unan}}(x) \\
\frac{1}{1 + e^{-u_{\text{Unan}}(x)}} &= \frac{1}{(1 + e^{-(\alpha_A + \beta_A x)})(1 + e^{- ( \alpha_B + \beta_B x )})} \\
\frac{1}{1 + e^{-u_{\text{Unan}}(x)}} &= \frac{1}{1 + e^{- \alpha_A - \beta_A x} + e^{- \alpha_B - \beta_B x} + e^{- \alpha_A - \alpha_B - \beta_A x - \beta_B x}} \\
e^{-u_{\text{Unan}}(x)} &= e^{- \alpha_A - \beta_A x} + e^{- \alpha_B - \beta_B x} + e^{- \alpha_A - \alpha_B - \beta_A x - \beta_B x} \\
-u_{\text{Unan}}(x) &= \log(e^{- \alpha_A - \beta_A x} + e^{- \alpha_B - \beta_B x} + e^{- \alpha_A - \alpha_B - \beta_A x - \beta_B x}).
\end{align*}

Thus, the utility function corresponding to unanimity is
\begin{equation}\label{eq:ut}
u_{\text{Unan}}(x) = - \log(e^{- \alpha_A - \beta_A x} + e^{- \alpha_B - \beta_B x} + e^{- \alpha_A - \alpha_B - \beta_A x - \beta_B x}).
\end{equation}
It is interesting to note that this is a complicated function, which does not look anything like the linear utility functions of the underlying individuals.

A challenge posed by this complicated functional form is that it is hard to interpret: it is not clear what this function does to the utility functions of $A$ and $B$.  However, this functional form---known as ``LogSumExp'' (a logarithm of a sum of exponentials)---has appeared in various literatures and is well understood (see, e.g., \citealt[\p 72]{Boyd04}, \citealt{Lugo20}).  The key insight is that this function closely approximates the minimum of its arguments, essentially tracing along whichever individual utility function is lowest at each point.  For a detailed explanation of this approximation and its bounds, see Appendix~\ref{app:logsumexp}.

Panel (a) in Figure~\ref{fig:unan-poly} illustrates this property, showing the utility function of unanimity (marked as a thick line) along with $u_A(x)$, $u_B(x)$, and $u_A(x) + u_B(x)$ for individuals with utility functions $u_A(x) = 5 + x$ and $u_B(x) = -5 + 3x$.  Notice how in each of the three segments of panel (a), separated by vertical lines, the organizational utility function traces the minima of the individuals' utility functions.\footnote{All figures plotting an organizational utility function use the actual utility function $u(x)$, not its approximation.  Given the tight approximation bounds (see Appendix~\ref{app:logsumexp}) and the plotted range, the actual function and its approximation look very similar.  For example, in Figure~\ref{fig:unan-poly}(a) $u_\text{Unan}(4)=6.87$ while $\min(u_A(4),u_B(4),u_A(4)+u_B(4))=u_B(4)=7$.}

\begin{figure}\centering
\includegraphics[width=\linewidth]{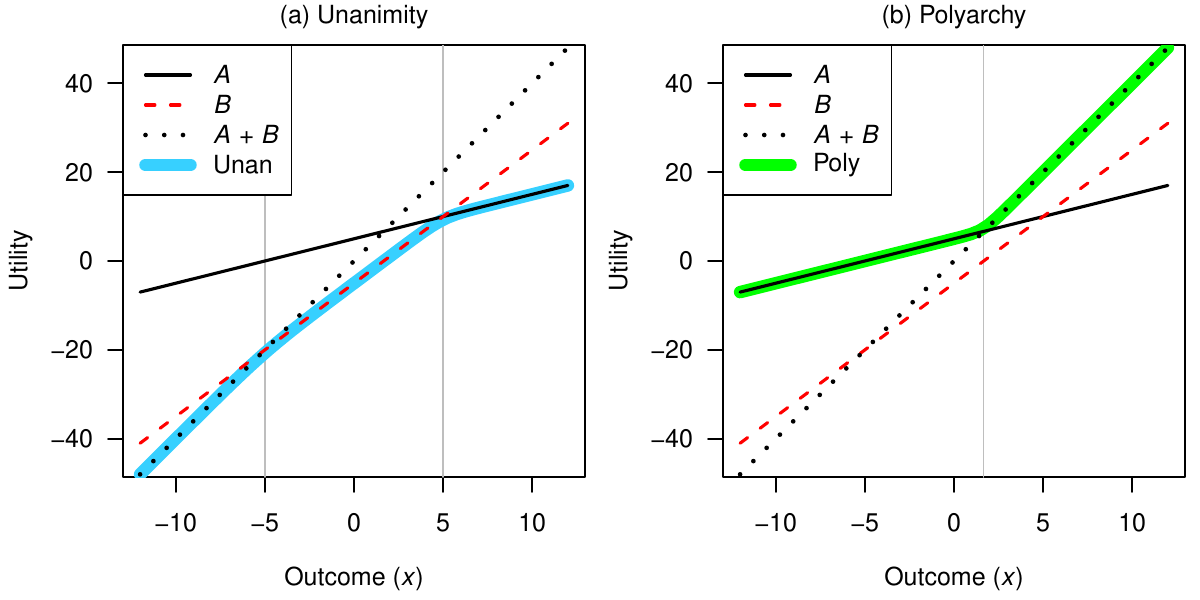}
\caption{Utility functions for (a) unanimity and (b) polyarchy.}\label{fig:unan-poly}
\end{figure}

The utility function of unanimity we have uncovered is surprising for three reasons.  First, it essentially corresponds to a simple mathematical function, the minimum of other utility functions.  Second, the utility function of unanimity not only depends on the utility of $A$ and the utility of $B$, but also on the utility of another individual that is not an actual member of the group, ``individual $A + B$,'' whose utility function is the sum of the utility functions of $A$ and $B$ (the implication of this will be explored in the Results section).  Finally, recall that marketing research had theorized that the utility function of a group was the average of the utility functions of its members.  However, this clearly does not hold true for unanimity, and it will not be the case for polyarchy either, as we will observe next.

\subsection{Polyarchy}

We now repeat the same steps as above to derive the utility function of the polyarchy.  The polyarchy approves a project when individual $A$ \textsc{or} $B$ believes the project is a good one.  Hence, the screening function of the polyarchy is:\footnote{The following derivations use the identity $1-1/(1+e^{-z}) = 1/(1+e^z)$.}
\begin{align*}
s_{\text{Poly}}(x) &= 1 - (1-s_A(x))(1-s_B(x)) \\
 &= 1 - \left( 1 - \frac{1}{1 + e^{-\alpha_A - \beta_A x}} \right) \left( 1 - \frac{1}{1 + e^{-\alpha_B - \beta_B x}} \right) \\
 &= 1 - \frac{1}{1 + e^{\alpha_A + \beta_A x}} \frac{1}{1 + e^{\alpha_B + \beta_B x}} \\
 &= 1 - \frac{1}{1 + e^{\alpha_A + \beta_A x} + e^{\alpha_B + \beta_B x} + e^{\alpha_A + \beta_A x + \alpha_B + \beta_B x}}.
\end{align*}

As before, we solve for the deterministic component of $u_{\text{Poly}}(x)$ that produces this screening function:
\begin{align*}
s(x) &= s_{\text{Poly}}(x) \\
\frac{1}{1 + e^{-u_{\text{Poly}}(x)}} &= 1 - \frac{1}{1 + e^{\alpha_A + \beta_A x} + e^{\alpha_B + \beta_B x} + e^{\alpha_A + \beta_A x + \alpha_B + \beta_B x}} \\
-1 + \frac{1}{1 + e^{-u_{\text{Poly}}(x)}} &= - \frac{1}{1 + e^{\alpha_A + \beta_A x} + e^{\alpha_B + \beta_B x} + e^{\alpha_A + \beta_A x + \alpha_B + \beta_B x}} \\
1 - \frac{1}{1 + e^{-u_{\text{Poly}}(x)}} &= \frac{1}{1 + e^{\alpha_A + \beta_A x} + e^{\alpha_B + \beta_B x} + e^{\alpha_A + \beta_A x + \alpha_B + \beta_B x}} \\
\frac{1}{1 + e^{u_{\text{Poly}}(x)}} &= \frac{1}{1 + e^{\alpha_A + \beta_A x} + e^{\alpha_B + \beta_B x} + e^{\alpha_A + \beta_A x + \alpha_B + \beta_B x}} \\
u_{\text{Poly}}(x) &= \log \left( e^{\alpha_A + \beta_A x} + e^{\alpha_B + \beta_B x} + e^{\alpha_A + \beta_A x + \alpha_B + \beta_B x} \right),
\end{align*}

In other words, polyarchy does the opposite of unanimity to members' utilities.  Panel (b) in Figure~\ref{fig:unan-poly} illustrates this.

To summarize, Figure~\ref{fig:functions-diagram} retraces our approach and the derivations in this section: (a) starting from individual utility functions, (b) we compute the corresponding individual acceptance probabilities, (c) we then aggregate these probabilities, and (d) finally ask what would be the utility function that would produce these aggregate choice probabilities.  The resulting utility function is the organizational utility function.  The arrows connecting these four elements correspond to going back and forth between utility and probability space (arrows 1 and 3) and aggregating probabilities (arrow~2).

\section{Analyses}\label{sec:analyses}

We now examine the behavior of the model under different scenarios.  The goal is to develop deeper intuitions about how the behavior of unanimity and polyarchy depends on the types of individuals employed.  We do so by developing four analyses: (i) comparing the risk-taking behavior of unanimity and polyarchy; (ii) increasing the number of organizational members; (iii) relaxing the ``more is better'' assumption to explore the effect of employing members with opposing interests (i.e., one utility function increases with $x$ while the other decreases with $x$); and (iv) examining how the results of classic strategic interaction models (Cournot competition and the principal--agent model) change when firms, instead of being risk-neutral, use a utility function corresponding to unanimity or polyarchy.  In Appendix~\ref{app:complex-utfns}, we illustrate the versatility of our approach by examining more complex individual utility functions.

\subsection{Comparing the Risk-Taking Behavior of Unanimity and Polyarchy}

So far we have shown that unanimity and polyarchy produce utility functions with very different shapes.  In comparison with its members, unanimity amplifies negative utilities and dampens positive ones, while polyarchy does the opposite.  These contrasting utility shapes have significant implications for risk assessment in project selection.  To quantify these differences, we employ the standard concept of ``certainty equivalent''---the guaranteed amount that would provide equal utility to a given risky proposition.  This approach allows us to systematically compare how unanimity and polyarchy structures evaluate and respond to projects with uncertain outcomes.

\begin{figure}\centering
\includegraphics[width=\linewidth]{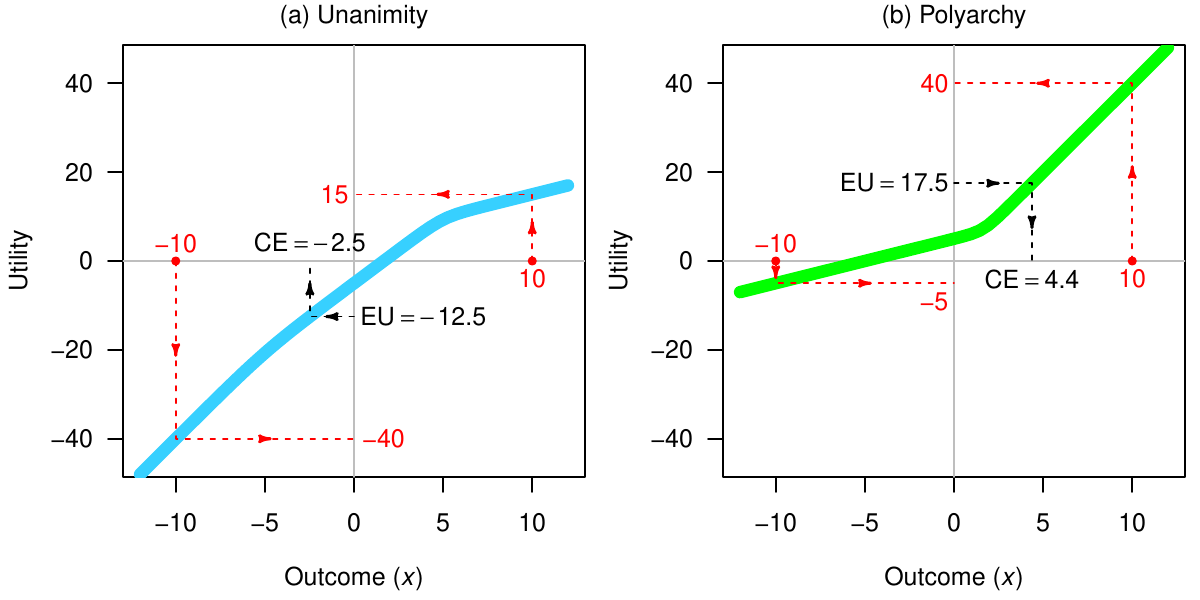}
\caption{Certainty equivalent (CE) of a bet that pays either $10$ or $-10$ with probability 50\% for unanimity and polyarchy.}\label{fig:cert-equivs}
\end{figure}

We use the same individual utility functions and aggregation schemes (unanimity and polyarchy) as in Figure~\ref{fig:unan-poly} to illustrate how aggregation can generate markedly different organizational risk preferences.  Specifically, the individuals' deterministic utility functions are $u_A(x)=5+x$ and $u_B(x)=-5+3x$ (with the usual $\mathrm{Logistic}(0,1)$ random-utility interpretation in the background, as in Section~\ref{sec:method}).  Figure~\ref{fig:cert-equivs} shows how unanimity and polyarchy evaluate a bet that pays either $10$ or $-10$ with equal probability.  Under unanimity, the bet yields expected utility $-12.5$ ($=0.5 u_\text{Unan}(-10) + 0.5 u_\text{Unan}(10) = 0.5 \times -40 + 0.5 \times 15$); under polyarchy, it yields expected utility $17.5$ ($=0.5 u_\text{Poly}(-10) + 0.5 u_\text{Poly}(10)=0.5 \times -5 + 0.5 \times 40$).  Using certainty equivalents, unanimity treats the bet as equivalent to a sure loss of 2.5 ($u_\text{Unan}(-2.5)=-12.5$), whereas polyarchy treats it as equivalent to a sure gain of 4.4 ($u_\text{Poly}(4.4)=17.5$).  The implication is that polyarchy is far more likely to accept this bet than unanimity.\footnote{We say ``most likely'' as these numerical examples did not incorporate the random utility component; however, the probability that $\mathrm{Logistic}(0,1)$ noise would flip the sign of either expected utility is negligible in the example.}

A further illustration of the substantial difference in risk-aversion between unanimity and polyarchy is to think about what would be the minimum winning probability at which they would accept the bet (i.e., $p$ such that $p \times u(10) + (1-p) \times u(-10) = 0$).  For the utility functions in Figure~\ref{fig:cert-equivs}, unanimity would only accept the bet if the probability of winning was at least 73\%, while polyarchy would just require an 11\% probability of winning.  In sum, aggregation structures have a paramount effect on risk-taking behavior.\footnote{While scaling the utility functions matters, it does not alter the fundamental finding that unanimity is more risk-averse and polyarchy more risk-seeking.  For instance, if the shape of the utility functions in Figure~\ref{fig:cert-equivs} remained the same but the $y$-axis range was compressed (e.g., from $-4$ to $+4$), polyarchy would still be more likely than unanimity to accept risky bets.  Of course, if the range were minimal (e.g., $-0.04$ to $+0.04$), noise would dominate, leading both structures to approve bets almost at random.}

The pronounced difference in risk aversion between unanimity and polyarchy, illustrated by their certainty equivalents for the same bet, highlights how aggregation structures shape organizational risk preferences derived from individual utilities.  This analytical insight can inform our understanding of strategic decisions involving risk, such as corporate acquisitions.  While studies often link acquisition choices to CEO attributes \citep{Zhu15} or incentives \citep{Wrig07}, major acquisitions typically involve top management teams with heterogeneous risk preferences.  Our framework suggests that aggregating these diverse preferences through structures like unanimity (more cautious) versus polyarchy (more risk-seeking) could help explain variations in acquisition aggressiveness, thus connecting individual managerial attributes to organizational outcomes \citep[\p 115]{Fink09}.

\subsection{Effect of Increasing the Number of Members in the Organization}\label{sec:larger-n}

Using our method, one can derive the organizational utility functions given arbitrary individual utility functions and aggregation structures.  Here we explore the case of unanimity and polyarchy as the number of members increases.\footnote{To compute the formulas, one can redo the analyses in Section~\ref{sec:method} using larger groups.  But a simpler way of doing so is to rely on the fact that \textsc{and} and \textsc{or} are associative functions.  For example, to derive the utility function of unanimity among three individuals ($A$, $B$, and $C$), one can compute the utility of $A$ \textsc{and} $B$ and call that $AB$, and then compute the utility of $AB$ \textsc{and} $C$.}

\begin{figure}\centering
\includegraphics[width=\linewidth]{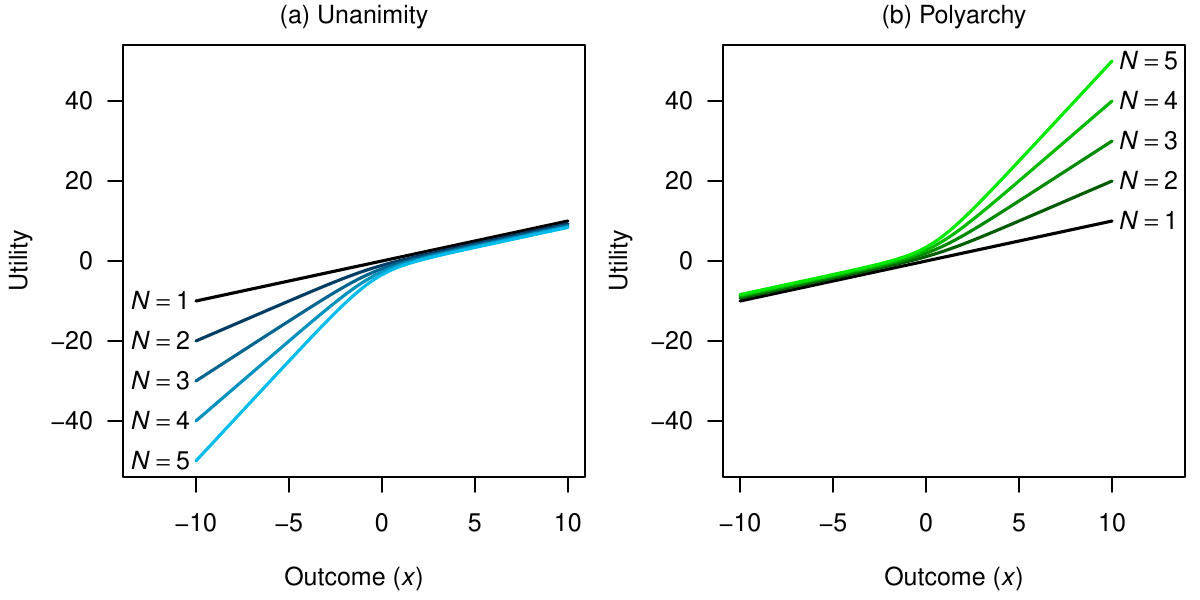}
\caption{Effect of varying the number of individuals ($N$) for unanimity and polyarchy.  The figure assumes all individuals have utility function $U_i(x) = x + \varepsilon_i$.}\label{fig:vary-n}
\end{figure}

Figure~\ref{fig:vary-n} illustrates the utility function of unanimity and polyarchy as the number of members ($N$) increases.  For simplicity, we assume that each member $i$ has utility function $U_i(x) = x + \varepsilon_i$.

The main observation from this figure is that, as $N$ increases, unanimity becomes more negative only for projects with negative $x$'s, while polyarchy becomes more positive only for projects with positive $x$'s.  This implies that unanimity becomes more risk-averse and polyarchy more risk-seeking as the number of members increases.  To understand what drives these results, it is useful to walk through the case of unanimity among three members (i.e., the $N=3$ line in Figure~\ref{fig:vary-n}(a)).  The key is to realize that unanimity among three individuals is the same as unanimity between one individual and a group of two individuals who use unanimity (i.e., aggregating the $N=1$ and $N=2$ lines in Figure~\ref{fig:vary-n}(a)).  Doing so with the $N=1$ and $N=2$ lines produces the $N=3$ line.  The same logic can be repeated for larger $N$'s as well as for polyarchy.

This analysis demonstrates how the derived organizational utility function, and consequently its risk preference, systematically shifts with team size ($N$) and aggregation structure.  Specifically, increasing $N$ magnifies risk aversion under unanimity and risk-seeking under polyarchy.  These predictable patterns offer clear avenues for empirically testing the link between team size, aggregation structure, and observed organizational risk-taking behavior.

\subsection{Effect of Opposing Views within the Organization}\label{sec:opposing-views}

So far we have studied individual utility functions that are increasing in $x$---although individuals may have different utility functions, they all agree that more of $x$ is better.  This is a reasonable assumption in many cases, such as when $x$ represents a generally desirable attribute (e.g., profits, sales, or ``likes'').  Individuals may disagree on the utility of a given project and on whether a given project is worth accepting, but they agree that increasing $x$ is a good thing.

\begin{figure}\centering
\includegraphics[width=\linewidth]{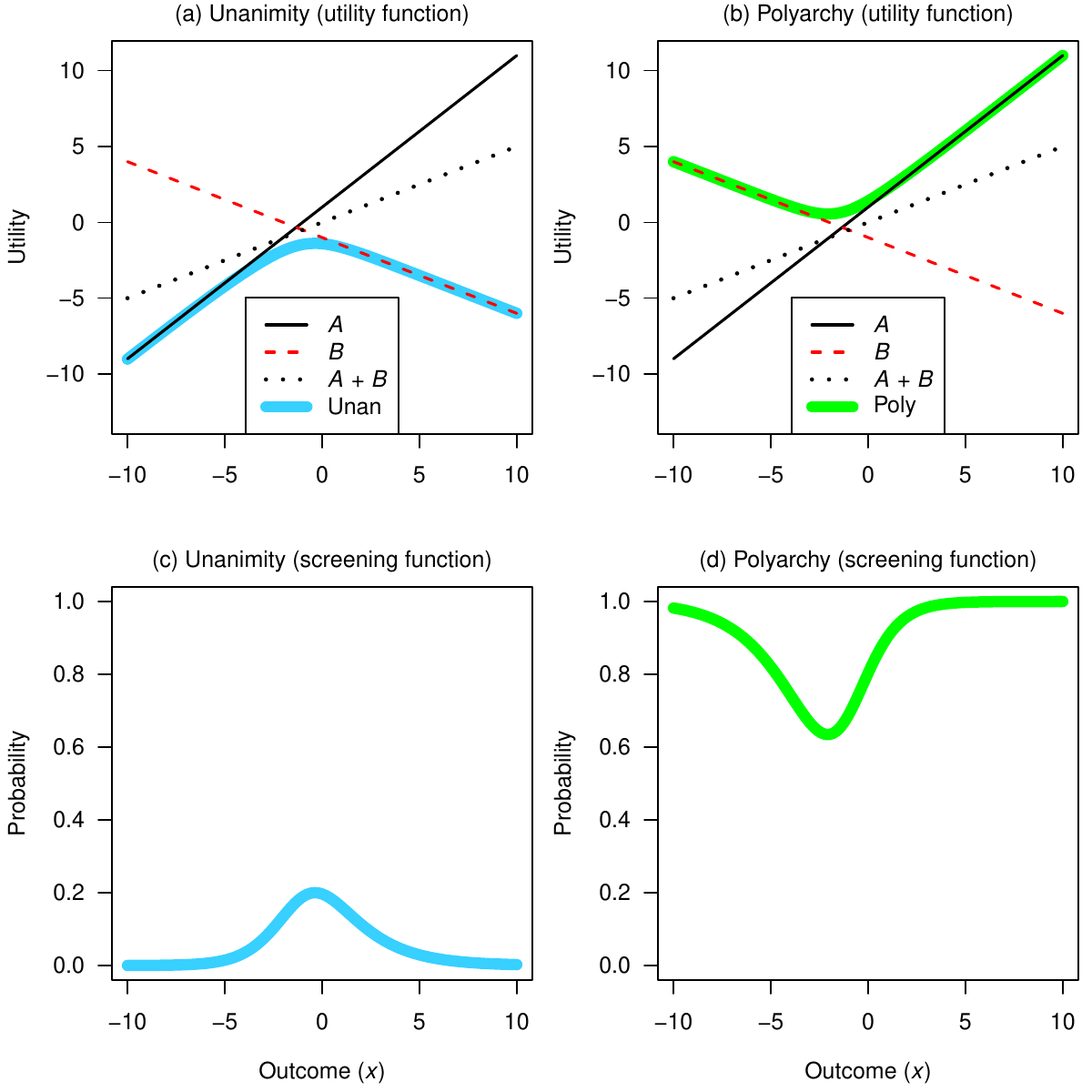}
\caption{Utility and screening functions of unanimity and polyarchy when the individuals have opposing views.  The thick lines in panels (a) and (b) show organizational utility functions and panels (c) and (d) show the respective organizational screening functions.}\label{fig:unan-poly-opposing}
\end{figure}

However, there are cases where the ``more is better'' assumption is not appropriate.  For example, if $x$ represents a trait of a possible hire or where to locate.  In cases like these, disagreements could run deep (e.g., members could disagree on whether to hire a younger or an older manager or on whether to locate close or far away from downtown).  Here we analyze the effect of using unanimity and polyarchy with two individuals, one who likes and another who dislikes increasing $x$.  Figure~\ref{fig:unan-poly-opposing} illustrates such a case (the individuals' utility functions used are $U_A(x) = 1 + x + \varepsilon_A$ and $U_B(x) = -1 - 0.5 x + \varepsilon_B$).  Again, we use simple linear utility functions for illustration; however, in this case, the slopes of the individuals are of a different sign to reflect opposite preferences.

Panel (a) in Figure~\ref{fig:unan-poly-opposing} shows the utility of unanimity under these managers with opposing views.  The organizational utility function in such a case follows an inverted-U shape.  As before, the organizational utility function traces the minima of the underlying utility functions.  In the case depicted in panel (a), this implies that on the negative side, the organizational utility function picks the upward-sloping utility function of $A$ and, on the positive side, the downward-sloping utility function of $B$.  Panel (b) illustrates the same individuals, now aggregated using polyarchy.  In this case, the organizational utility function is now U-shaped.

Panels (c) and (d) in Figure~\ref{fig:unan-poly-opposing} show the screening functions corresponding to the organizational utility functions from panels (a) and (b), respectively.  An interesting observation here is that these screening functions look very different than the S-shaped screening functions depicted earlier in this paper (Figures \ref{fig:ex-screening} and \ref{fig:ex-utility-screening}).  One can understand this qualitative difference by following how the utility function affects the screening function.  For an increasing utility function that goes from very negative to very positive utilities, its screening function will shift from mostly rejecting to mostly approving projects and, hence, will produce an S-shaped screening function like the ones in Figure~\ref{fig:ex-screening}.  In contrast, if a utility function follows an inverted U-shape (as in panel (a) in Figure~\ref{fig:unan-poly-opposing}), its corresponding screening function will be most likely to accept projects at the apex of the inverted U-shape, which produces a bell-shaped screening function like the one shown in panel (c).  The opposite happens for a polyarchy employing opposing individuals, which produces a U-shaped utility function and whose screening function achieves a minimum coinciding with the U-shape's nadir (i.e., panels (b) and (d) in Figure~\ref{fig:unan-poly-opposing}).

Panel (c) sheds light on situations characterized by deep disagreements.  For instance, one could imagine this panel representing an academic department with two factions: a group that likes hiring theoreticians and a group that likes hiring empiricists (or in a corporate setting, this could represent a top management team divided between those prioritizing operations and those prioritizing marketing).  The $x$-axis in this case represents the spectrum from pure theoreticians to pure empiricists.  If hiring required unanimity, such a department would act as if its utility function followed an inverted U-shape.  That is, only candidates who are neither too theoretical nor too empirical would be somewhat appealing to the department (as in panel (a)).  Hence, this department would only accept candidates close to the middle of the spectrum.  Moreover, because no candidate appeals across the board, the organizational utility is never very high and, consequently, this department may not hire very often (e.g., the screening function in panel (c) approves the most appealing candidate around 20\% of the time).  Of course, the particulars of conflict for different organizations and decisions will be different from those illustrated in Figure~\ref{fig:unan-poly-opposing}, but considering how an organization's utility function may look can help predict what the most likely organizational outcomes might be.

This formal approach to aggregation sheds light on how organizations manage the internal conflict central to the Behavioral Theory of the Firm \citep{Cyer63}.  Our model demonstrates \emph{how} conflicting preferences among individuals or coalitions can be reconciled---or ``quasi-resolved''---into organizational action based on the aggregation structure.  As the preceding analysis of opposing views illustrates (Figure~\ref{fig:unan-poly-opposing}), unanimity forces compromise, favoring outcomes only where conflicting preferences find common ground (akin to a negotiated truce).  Conversely, polyarchy allows the preferences of any single actor favoring action to drive the outcome, even amidst dissent.  This provides a clear mechanism for understanding how seemingly coherent organizational goals can emerge despite, and indeed \emph{as a result of}, aggregating conflicting internal viewpoints.

\subsection{Organizational Utility in Strategic Interactions: Applications to Collaboration and Competition}\label{sec:strat-interactions}

We now illustrate the strategic implications of using these derived organizational utility functions by applying them to two classic interaction models, substituting the typical risk-neutral assumption with the distinct utility functions for unanimity and polyarchy developed earlier.  First, we examine a Cournot model, analyzing how firms' internal aggregation structures shape competitive output decisions.  Second, we analyze a principal--agent model, representative of collaborative relationships like firm--supplier contracting, exploring how the principal's aggregation structure influences optimal incentive design.  These applications highlight how replacing the simplistic risk-neutral unitary actor assumption with behaviorally grounded organizational utility functions leads to richer predictions about strategic behavior in both competitive and collaborative settings (for complete technical derivations and parameter specifications, see Appendix~\ref{app:details}).

\begin{figure}\centering
\includegraphics[width=\linewidth]{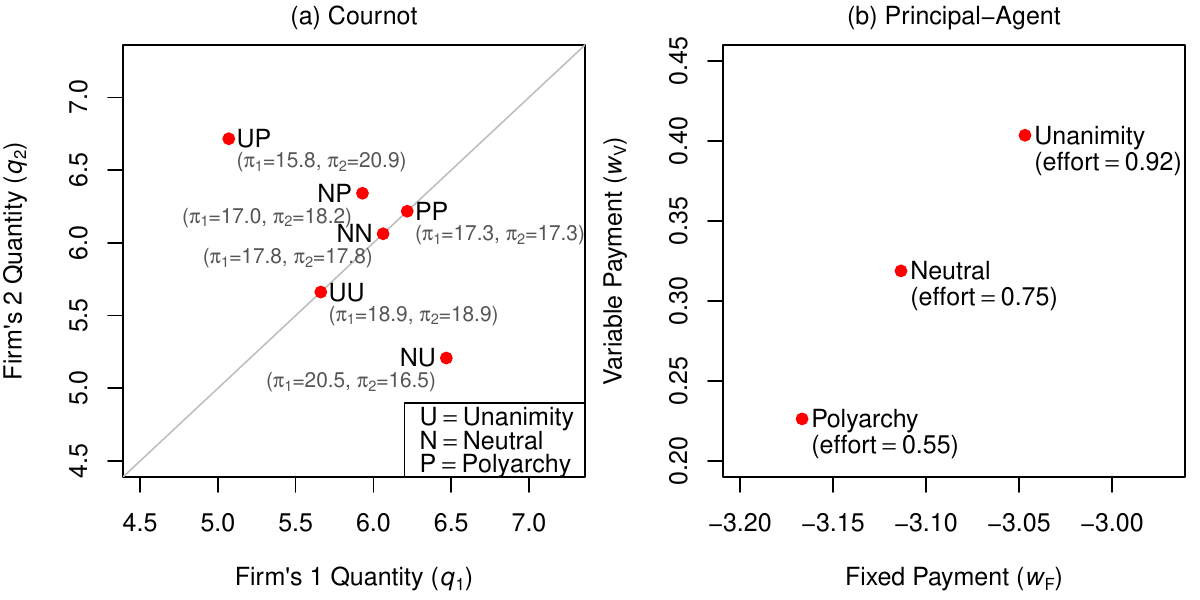}
\caption{Equilibrium outcomes under different decision-making structures.  (a) Production levels in Cournot competition between firms with different decision-making structures.  (b) Principal--agent contracts and resulting effort levels under different principal decision-making structures.}\label{fig:io}
\end{figure}

First, consider a Cournot duopoly where firms simultaneously choose production quantities under demand uncertainty.  We compare all possible competitive scenarios among unanimity, polyarchy, and risk-neutral firms (labeled as $U$, $P$, and $N$ in Figure~\ref{fig:io}(a)).  The game features a demand function with uncertainty in its intercept.\footnote{In this example, $\mathit{demand} = \mathit{intercept} - 0.5(q_1+q_2)$, where $\mathit{intercept} \sim \mathrm{Normal}(10,2^2)$.}  As Figure~\ref{fig:io}(a) shows, production is highest when two polyarchies compete (point $PP$), because polyarchy's risk-seeking nature sees its utility increase the most with increased production.  This implies that polyarchical firms in oligopolistic markets will tend to overproduce, whereas unanimity-based firms will tend to produce less.  Interestingly, while firms would achieve higher profits by being more risk-averse and restricting production (point $UU$), competitive pressures could drive them toward more risk-seeking structures, which increase output but reduce profitability.

Second, we examine a classic principal--agent relationship, modeling, for instance, a firm contracting with a supplier or employee.  We assume a standard risk-averse agent \footnote{We use a CARA utility function $u_A = -\exp(-0.5 \cdot \mathit{wage}) - \tfrac{1}{2}\mathit{effort}^2$ and reservation utility of $-5$, and assume $\mathit{output} \sim \mathrm{Normal}(\mathit{effort},3^2)$.}  The principal sets a linear contract ($\mathit{wage} = w_\text{F} + w_\text{V} \mathit{output}$).  Figure~\ref{fig:io}(b) shows that a unanimity principal sets the strongest incentives (higher fixed and variable payments), thereby inducing maximum effort.  This occurs because more risk-averse principals experience greater utility loss from low outputs and consequently set stronger incentives to prevent such outcomes.  Less risk-averse principals do not prioritize such high effort as they are unconcerned with low output.  Ultimately, risk-averse unanimities will tend to pay their agents (e.g., suppliers) higher rates to avoid the risk of low production.  In contrast, risk-seeking polyarchies will tend to pay their agents less in order to optimize profitability while accepting the risk of reduced production.

These examples show how replacing the risk-neutral unitary actor assumption with realistic organizational utility functions offers different testable predictions about how individual preferences and aggregation structures translate into observable strategic behavior.

\section{Discussion}

Traditional strategy theories often depict firms as unitary actors, simplifying the complex interplay of individual preferences within an organization and assuming that the whole firm can be represented as having one utility function.  The behavioral theory of the firm and the literature on microfoundations highlight that the unitary actor is an illusion: firms are composed of multiple actors with oftentimes conflicting goals, suggesting that the whole firm should not be understood as a unitary actor.

However, we show that if one knows the utility functions of the members of the organization and the aggregation structure they use, it is possible to derive an organizational utility function---one utility function that accurately represents the preferences of the organization.  Thus, although the organization is not a unitary actor, if one uses the right organizational utility function, it is possible to analyze the organization \emph{as if} it was a unitary actor.  So, while the firm may be a motley crew behind the scenes, it can still put on a convincing one-person show for observers.

This approach bridges traditional strategy theories with a deeper understanding of organizational behavior.  It acknowledges the complexity of internal dynamics while offering a practical analytical framework.  By recognizing both organizational intricacy and the potential for unified representation, we pave the way for more accurate and insightful strategy research.  The following discussion explores the implications of our findings.

\subsection{Contributions to the Strategy Literature}

First, we show that organizational utility functions generally differ from the average of their members' utility functions, and the type of aggregation structure used greatly affects the shape of the resulting organizational utility function.\footnote{Because the space of possible aggregation structures is vast (see, e.g., \citealt{Csas13,Csas13b}), it is likely that some aggregation structure could effectively average its members' utility functions---what we have pointed out is that this is not generally the case.}  Specifically, unanimity traces the minima of underlying utility functions, while polyarchy traces their maxima.  Interestingly, these underlying utility functions are not just the utility functions of the individual members of the organization but also include the utility functions of other, ``synthetic'' members that often amplify the characteristics of the actual members of the organization.

Second, our work provides a theoretical basis for revisiting unitary actor models to incorporate the role of individual heterogeneity and aggregation structures.  We show that it is valid to conceptualize an organization in terms of a utility function as long as such a function is chosen carefully.  This allows existing theories to be ``retrofitted'' by replacing their utility functions with ones that better reflect the characteristics of aggregation processes.  For instance, search and industrial organization models commonly used in strategy assume risk-neutral firms (in terms of our model, such research assumes the organization's utility function is a straight line; \citealt[\p 631]{Roth76}).  However, our work shows that aggregation can exacerbate the characteristics of underlying utility functions; thus, most organizational utility functions, which arise from aggregation processes, may look quite different from those associated with unitary actor models so far.  Retrofitting previous unitary actor theories offers a straightforward path for incorporating the role of individual heterogeneity and aggregation structures into these theories.

Third, we provide a way to address the call to better incorporate conflict into the behavioral theory of the firm \citep{Gave07}.  Our framework demonstrates how specific aggregation structures systematically combine potentially conflicting individual or coalition preferences (see, e.g., the predictions in the context of Figure~\ref{fig:unan-poly-opposing}).  This process yields a distinct organizational utility function, offering a path to reconcile the apparent contradiction between viewing the organization as a coalition of diverse interests and ascribing it coherent goals \citep[\p 27]{Cyer63}.  Moreover, by deriving this function analytically, our approach provides a tractable method for representing internal conflict's impact on organizational preferences, tackling a long-standing challenge \citep[\pp 668--671]{Marc62}.

Finally, our work can help the top management teams (TMT) literature establish a better link between the causes and effects it studies.  \citet[\p 115]{Fink09} observation that ``the gulf between executive characteristics and organizational outcomes is huge'' is a call to better understand the chain of causality studied by this literature, which spans from individual characteristics (e.g., narcissism and functional background) to organizational outcomes (e.g., R\&D investments and profitability).  If such a chain of causality can be expressed in terms of individual utility functions, aggregation mechanisms, and organizational utility functions---and we think it can, as TMTs operate at the intersection of these elements---our work can help bridge the gulf at the core of the TMT literature.  Hence, our work also suggests new, testable predictions for this literature regarding the effect of aggregation structures.

\subsection{Limitations and Further Work}

As with all research, our work has limitations, which provide opportunities for future research.  We identify four main areas for extension.

First, our analysis used stylized aggregation structures (unanimity, polyarchy) and simple individual utility functions.  Future work could incorporate a wider array of aggregation mechanisms, such as majority voting, delegation, or weighted schemes representing power differentials (e.g., giving veto power or extra weight to certain individuals like a dictatorial CEO; \citealt{Csas13,Csas13b}).  Further extensions could model more complex individual utilities, perhaps incorporating multiple attributes, and explicitly distinguish between conflicts among individuals \citep{Eise97} and conflicts among goals \citep{Gaba19,Oblo20}, potentially modeling the latter via different ways of aggregating multi-attribute utility functions.

Second, our model is static.  Incorporating dynamics---how individual utilities or aggregation structures evolve through learning \citep{Piez22}, strategic adaptation \citep{Piez23}, or other decision processes---would provide richer insights into organizational change and adaptation.  This could help illuminate phenomena like shifts in TMT decision patterns \citep{Fink09} and incumbent responses to technological change \citep{Egge18}; for example, how the interplay between management team composition, individual utilities, and the aggregation mechanism shapes the firm's overall risk preference, influencing decisions about whether to invest in or actively capture value from emerging technologies \citep{Chri96}.

Third, regarding empirical work, future research could test our model in the lab by eliciting the utility functions of group members and checking whether the utility function of the group matches our predictions.  Apart from validating our work, this research would shed light on the effect size of aggregation structure vis-à-vis other individual-, firm-, and environmental-level contingencies.

Finally, our framework naturally extends to hybrid decision-making bodies where humans and AI agents jointly evaluate opportunities.  By treating an AI system as another organizational ``member'' with its own utility function and error profile, researchers can analyze human--AI collaboration using the same primitives as human-only teams.  Like human members, an AI's utility function is defined by its training and constraints, meaning it may be misaligned with organizational goals or opaque to designers.  Consequently, firms must validate what the AI is actually optimizing behaviorally rather than assuming it is a neutral oracle.  The key design implication is that firms must jointly choose the AI's utility function and its structural role---such as a gatekeeper (unanimity) or a spotter (polyarchy)---as these choices fundamentally reshape the organization's risk profile.  This perspective opens a new research agenda focused on characterizing AI utility functions and designing aggregation structures that optimally aggregate human and machine recommendations.

These extensions would improve our understanding of how micro-level preferences and power structures translate into macro-level organizational behavior and strategic decision-making.

\subsection{Conclusion}

Our paper provides insight into how heterogeneous individual utility functions are aggregated by different decision-making structures.  In doing so, we demonstrate that organizational utility functions are a meaningful construct and that they are not just an average of individuals' utility functions.  Rather, part of an organization's utility function is directly inherited from its members' utility functions, but part of it also comes from adding up individuals' utility functions.  How these parts are combined critically depends on the aggregation structure used by the organization.  The picture that emerges is that organizations can exaggerate the behavioral characteristics of their members.  It is the role of the organization designer to harness these exaggerated traits in ways that benefit the organization.  Ultimately, organizational utility functions can help provide keener insights into how organizations make the decisions they do.  Overall, our goal has been to logically connect key micro and macro aspects of organizations and, thus, to help provide a sound foundation to understand the behavior of organizations based on the attributes of their underlying individuals.

\clearpage
\begin{singlespacing}
\bibliography{refs}
\end{singlespacing}

\clearpage
\appendix
\FloatBarrier
\section{The LogSumExp Approximation}\label{app:logsumexp}

This appendix provides a detailed explanation of the LogSumExp approximation, which offers a convenient intuition for interpreting the organizational utility functions derived in the main text.

\subsection{The LogSumExp Function and Its Bounds}

The LogSumExp function is defined as $\text{LSE}(x_1, \ldots, x_n) = \log(e^{x_1} + e^{x_2} + \ldots + e^{x_n})$.  This function appears frequently in machine learning, convex optimization, and statistical mechanics (see, e.g., \citealt[\p 72]{Boyd04}, \citealt{Lugo20}).  A key property of this function is that it closely approximates the maximum function, with the following tight bounds:
\[
\max(x_1, x_2, \ldots, x_n) \leq \log(e^{x_1} + e^{x_2} + \ldots + e^{x_n}) \leq \max(x_1, x_2, \ldots, x_n) + \log n
\]

\subsubsection{Proof}

Let $M = \max(x_1, x_2, \ldots, x_n)$.

\paragraph{Lower Bound.}  Since $M$ is the maximum, $e^M$ is the largest term in the sum $\sum_{i=1}^{n} e^{x_i}$.  Thus, $\sum_{i=1}^{n} e^{x_i} \geq e^M$.  Taking the logarithm (a monotonically increasing function) preserves the inequality:
\[
\log\left(\sum_{i=1}^{n} e^{x_i}\right) \geq \log(e^M) = M.
\]

\paragraph{Upper Bound.}  For all $i$, we have $x_i \leq M$, so $e^{x_i} \leq e^M$.  Summing over all $i$:
\[
\sum_{i=1}^{n} e^{x_i} \leq \sum_{i=1}^{n} e^M = n e^M.
\]
Taking the logarithm:
\[
\log\left(\sum_{i=1}^{n} e^{x_i}\right) \leq \log(n e^M) = \log n + M.
\]

\paragraph{Conclusion.}  Combining both bounds:
\[
M \leq \log\left(\sum_{i=1}^{n} e^{x_i}\right) \leq M + \log n.
\]

This shows that the LogSumExp function is tightly approximated by $\max(x_1, x_2, \ldots, x_n)$.  The approximation error is bounded above by $\log n$, which equals approximately 1.1 for $n = 3$ (the number of terms in our two-member organizational utility functions).  When the differences between the $x_i$ values are large, the approximation becomes even tighter, as the sum becomes dominated by the largest exponential term.

\subsection{Application to Organizational Utility Functions}

\subsubsection{Unanimity}

The utility function for unanimity derived in the main text is:
\[
u_{\text{Unan}}(x) = -\log(e^{-\alpha_A - \beta_A x} + e^{-\alpha_B - \beta_B x} + e^{-\alpha_A - \alpha_B - \beta_A x - \beta_B x}).
\]

This can be rewritten as $u_{\text{Unan}}(x) = -\text{LSE}(-\alpha_A - \beta_A x, \ -\alpha_B - \beta_B x, \ -\alpha_A - \alpha_B - \beta_A x - \beta_B x)$.

Applying the LogSumExp approximation:
\begin{align*}
u_{\text{Unan}}(x) &\approx -\max(-\alpha_A - \beta_A x, \ -\alpha_B - \beta_B x, \ -\alpha_A - \alpha_B - \beta_A x - \beta_B x) \\
&= \min(\alpha_A + \beta_A x, \ \alpha_B + \beta_B x, \ \alpha_A + \alpha_B + \beta_A x + \beta_B x) \\
&= \min(u_A(x), \ u_B(x), \ u_A(x) + u_B(x)).
\end{align*}

Thus, the utility function of unanimity traces the pointwise minimum of the individual utility functions and their sum.

\subsubsection{Polyarchy}

The utility function for polyarchy derived in the main text is:
\[
u_{\text{Poly}}(x) = \log(e^{\alpha_A + \beta_A x} + e^{\alpha_B + \beta_B x} + e^{\alpha_A + \alpha_B + \beta_A x + \beta_B x}).
\]

This is directly in LogSumExp form: $u_{\text{Poly}}(x) = \text{LSE}(\alpha_A + \beta_A x, \ \alpha_B + \beta_B x, \ \alpha_A + \alpha_B + \beta_A x + \beta_B x)$.

Applying the approximation:
\begin{align*}
u_{\text{Poly}}(x) &\approx \max(\alpha_A + \beta_A x, \ \alpha_B + \beta_B x, \ \alpha_A + \alpha_B + \beta_A x + \beta_B x) \\
&= \max(u_A(x), \ u_B(x), \ u_A(x) + u_B(x)).
\end{align*}

Thus, the utility function of polyarchy traces the pointwise maximum of the individual utility functions and their sum.

\section{More Complex Utility Functions}\label{app:complex-utfns}

For the sake of clarity, so far we have used linear individual utility functions that depend on just one parameter ($x$).  But neither of these assumptions is required for our method to work: the steps described in Figure~\ref{fig:functions-diagram} can be used to aggregate arbitrarily complex utility functions.  Below we illustrate this point by aggregating non-linear and multi-parameter utility functions.  As in the main text, utilities are interpreted in random-utility form (i.e., $U(\cdot)=u(\cdot)+\varepsilon$ with $\varepsilon \sim \mathrm{Logistic}(0,1)$); the figures below plot the deterministic components $u(\cdot)$.

\begin{figure}\centering
\includegraphics[width=\linewidth]{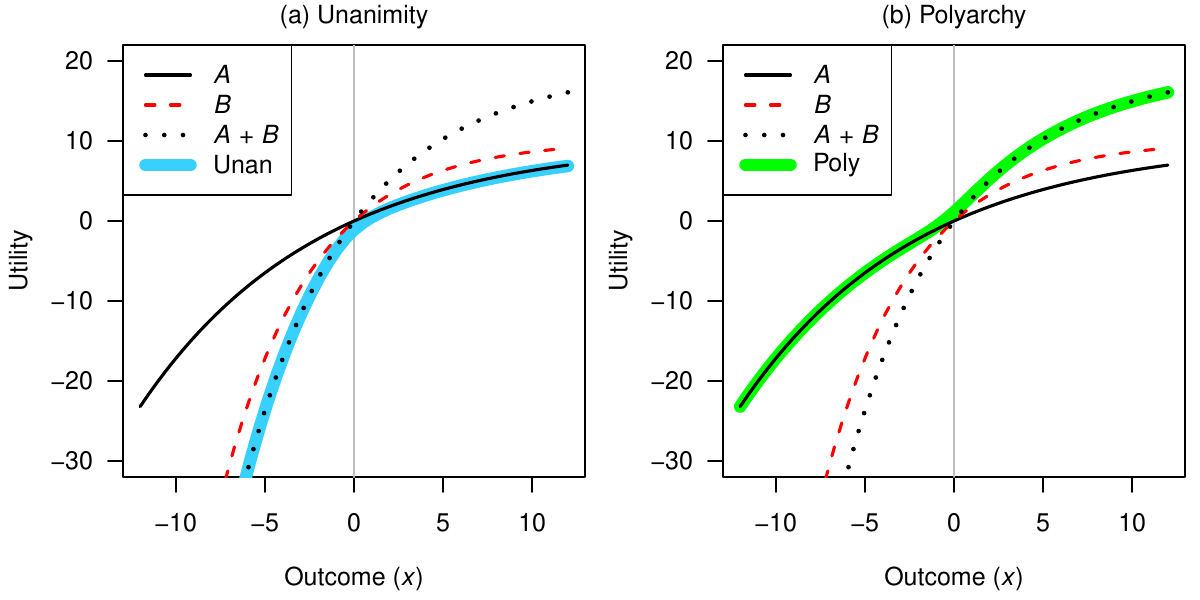}
\caption{Aggregating risk-averse individuals.}\label{fig:unan-poly-curves}
\end{figure}

Figure~\ref{fig:unan-poly-curves} illustrates how unanimity and polyarchy aggregate two risk-averse individuals.  The individuals in this example have exponential utility functions $u_A(x) = 10 (1 - e^{-x/10})$ and $u_B(x) = 10 (1 - e^{-x/5})$ and hence, exhibit constant absolute risk aversion.  Aggregation operates the same as before: unanimity and polyarchy trace the pointwise minima and maxima of the utility functions of $A$, $B$, and $A + B$.  As before also, unanimity exacerbates risk-aversion and polyarchy, risk-seeking.

\begin{figure}\centering
\includegraphics[width=\linewidth]{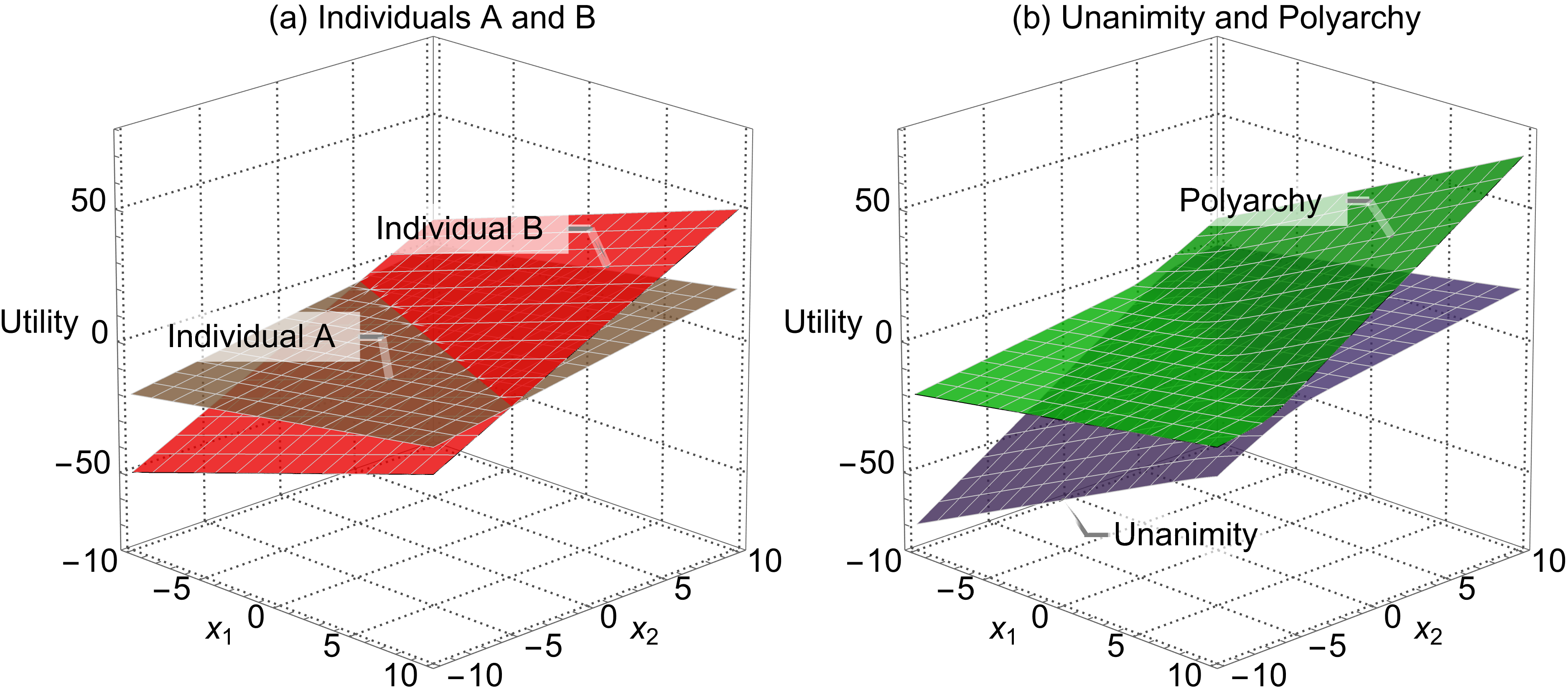}
\caption{Aggregating utility functions that depend on two arguments ($x_1$ and $x_2$).}\label{fig:utility-3d}
\end{figure}

Figure~\ref{fig:utility-3d} illustrates the aggregation of utility functions that depend on two arguments ($x_1$ and $x_2$).  For instance, these two variables could correspond to two characteristics of a project under consideration (e.g., the range and power of a car).  In the example, the two individuals differ in how much they value each characteristic.  Panel (a) plots the utility functions of both individuals ($u_A(x_1, x_2) = x_1 + x_2$ and $u_B(x_1, x_2) = 2 x_1 + 3 x_2$).  Panel (b) shows the corresponding utility function of unanimity and polyarchy.  Similarly to the previous results, the utility for polyarchy is always above that for unanimity.  Also, the organizational utility rises sharply for positive values of $x_1$ and $x_2$ for polyarchy and drops steeply for negative values of $x_1$ and $x_2$ for unanimity.

\section{Details of Strategic Interaction Analyses}\label{app:details}

This appendix provides the technical details, functional forms, and parameter values required to reproduce the analyses presented in Section~\ref{sec:strat-interactions}.  Throughout this appendix, we model organizations composed of two members, $A$ and $B$, with the same random-utility specifications used in Figure~\ref{fig:unan-poly}.  Specifically, $U_A(x) = 5 + x + \varepsilon_A$ and $U_B(x) = -5 + 3x + \varepsilon_B$, where the error terms $\varepsilon_A$ and $\varepsilon_B$ are independent draws from a $\mathrm{Logistic}(0,1)$ distribution.

\subsection{Cournot Competition Model}

We model a duopoly where two firms ($i$ and $j$) simultaneously choose production quantities ($q_i$ and $q_j$) to maximize their expected utility.

\emph{Market Structure.}  We assume a standard linear inverse demand function with uncertainty in the intercept, where market price $P$ depends on total quantity $Q = q_i + q_j$:
\[ P(Q) = \max(a - b(q_i + q_j), 0) \]
where $a$ is a random variable drawn from a Normal distribution $a \sim \mathrm{Normal}(\bar{a}, \sigma_a^2)$.  The marginal cost of production, $c$, is constant and identical for both firms.  The profit for firm $i$ is given by $\pi_i = (P(Q) - c)q_i$.

\emph{Firm Preferences.}  Firms evaluate profits according to their organizational utility function, $u(\pi)$.  We compare three types of firms: risk-neutral (unitary actor), unanimity, and polyarchy.  The risk-neutral firm employs a linear utility function $u(\pi) = \pi$.  The unanimity and polyarchy firms aggregate the preferences of two members ($A$ and $B$) using the functional forms derived in Section~\ref{sec:method}.

\emph{Optimization and Equilibrium.}  Each firm $i$ chooses $q_i$ to maximize its expected utility given the quantity produced by the competitor $q_j$.  We solve for the Nash Equilibrium numerically using an iterative best-response algorithm where firms update their quantities sequentially until convergence (defined as a change in quantity of less than $10^{-7}$ between iterations).  Expected utility integrals are computed numerically over the range $[\bar{a} - 10\sigma_a, \bar{a} + 10\sigma_a]$.

\emph{Parameters.}  Table~\ref{tbl:cournot-parameters} lists the market parameter values used to generate Figure~\ref{fig:io}(a).

\begin{table}[h]
\centering
\begin{tabular}{lcc}
\hline
\textbf{Parameter} & \textbf{Symbol} & \textbf{Value} \\ \hline
Mean demand intercept & $\bar{a}$ & $10$ \\
SD of demand intercept & $\sigma_a$ & $2$ \\
Demand slope & $b$ & $0.5$ \\
Marginal cost & $c$ & $1$ \\ \hline
\end{tabular}
\caption{Parameters for Cournot competition analysis.}
\label{tbl:cournot-parameters}
\end{table}

\subsection{Principal--Agent Collaboration Model}

We model a collaboration where a principal contracts a risk-averse Agent to perform a task.

\emph{Production and Contract.}  The output $R$ is determined by the Agent's effort $e$ and a random shock, such that $R \sim \mathrm{Normal}(e, \sigma^2)$.  The principal offers a linear contract consisting of a fixed wage $w_F$ and a variable component $w_V$ based on output, $w = w_F + w_V R$.

\emph{Agent Preferences.}  The Agent is risk-averse with Constant Absolute Risk Aversion (CARA) utility and a convex cost of effort.  The Agent's utility function is:
\[ U_{\text{Agent}}(w, e) = -e^{-\gamma w} - \frac{1}{2}e^2 \]
where $\gamma$ is the coefficient of absolute risk aversion.

\emph{Principal Preferences.}  The principal evaluates their net income ($R - w$) using one of the three organizational utility functions defined previously ($u_{\text{Neutral}}$, $u_{\text{Unan}}$, or $u_{\text{Poly}}$).

\emph{Optimization Problem.}  The principal chooses the contract terms ($w_F, w_V$) and the induced effort level $e$ to maximize their expected utility, subject to the Agent's Incentive Compatibility (IC) and Participation Constraints (PC):
\[
\begin{aligned}
\max_{w_F, w_V, e} \quad & \E_R \big[ u_{\text{Principal}}(R - w) \big] \\
\text{s.t.} \quad
& e = \arg\max_{e'} \, \E_R \big[ U_{\text{Agent}}(w, e') \big] && \text{(IC)} \\
& \E_R \big[ U_{\text{Agent}}(w, e) \big] \ge \underline{U} && \text{(PC)}
\end{aligned}
\]

The IC constraint is implemented via the Agent's first-order condition with respect to effort.  The problem is solved using numerical optimization (Simulated Annealing method).

\emph{Parameters.}  Table~\ref{tbl:pa-parameters} lists the task environment parameter values used to generate Figure \ref{fig:io}(b).

\begin{table}[h]
\centering
\begin{tabular}{lcc}
\hline
\textbf{Parameter} & \textbf{Symbol} & \textbf{Value} \\ \hline
Output noise (SD) & $\sigma$ & $3$ \\
Agent risk aversion & $\gamma$ & $0.5$ \\
Agent reservation utility & $\underline{U}$ & $-5$ \\ \hline
\end{tabular}
\caption{Parameters for principal--agent analysis.}
\label{tbl:pa-parameters}
\end{table}

\end{document}